\documentclass[journal,12pt,onecolumn]{IEEEtran}
\usepackage{setspace}
\usepackage{tikz}
\usetikzlibrary{shapes,arrows,fit,calc,positioning,automata}

\usepackage{cite}
\usepackage{hyperref}

\usepackage{array}
\usepackage{caption}

\usepackage{mdwmath}
\usepackage{mdwtab}
\usepackage{mathtools}
\usepackage{amsmath} 
\usepackage{amssymb}

\begin{document}

\title{Convolutional Network-Coded Cooperation in Multi-Source Networks with a Multi-Antenna Relay}

\author{Alireza Karbalay-Ghareh,
         Masoumeh Nasiri-Kenari,~\IEEEmembership{Senior Member,~IEEE}, and\\ Mohsen Hejazi
\thanks{The authors are with the Wireless Research Laboratory (WRL), Department
of Electrical Engineering, Sharif University of Technology, Tehran,
Iran (email: $\lbrace$karbalayghareh, mhejazi$\rbrace$ @ee.sharif.edu, mnasiri@sharif.edu).}}


\maketitle

\begin{abstract}
We propose a novel cooperative transmission scheme called \textit{Convolutional Network-Coded Cooperation} (CNCC) for a network including $N$ sources, one $M$-antenna relay, and one common destination. The source-relay (S-R) channels are assumed to be Nakagami-$m$ fading, while the source-destination (S-D) and the relay-destination (R-D) channels are considered Rayleigh fading. The CNCC scheme exploits the generator matrix of a good $(N+M^{'}, N, \nu)$ systematic convolutional code, with the free distance of $d_{free}$ designed over $GF(2)$, as the network coding matrix which is run by the network's nodes, such that the systematic symbols are directly transmitted from the sources, and the parity symbols are sent by the best antenna of the relay. An upper bound on the BER of the sources, and consequently, the achieved diversity orders are obtained. The numerical results indicate that the CNCC scheme outperforms the other  cooperative schemes considered, in terms of the diversity order and the network throughput. The simulation results confirm the accuracy of the theoretical analysis.
\end{abstract}


\section{Introduction}
\IEEEPARstart{o}{ne} of the most important and intrinsic features of the wireless networks is fading. This phenomenon induces many adverse effects in the networks, and considerably reduces the performance. Diversity is a well-known technique to deal with fading, which is used in the various domains such as time, frequency, and space. Cooperative relay-based networks have been proposed to combat fading, by benefiting from the spatial diversity through the relays or the antennas of a multi-antenna relay \cite{1,2,3,4}. The traditional cooperative transmission schemes, such as Amplify-and-Forward (AF) and Decode-and-Forward (DF), have been introduced and evaluated in the the numerous papers like \cite{4,5,6,7,8}. The basic and common shortcoming of these schemes is the reduction of the network throughput in the multi-sources networks \cite{9, 10}. To eliminate this problem, the idea of using network coding in the cooperative networks has been suggested in the recent years.

Network Coding was first introduced by Ahlswede et al. in the seminal work \cite{11}, and then proposed as Linear Network Coding (LNC) in \cite{12}. The main idea of the LNC is that in a multi-hop network, the intermediate nodes combine linearly the received data from the source nodes, and transmit them to the next hops, instead of transmitting each received data separately. This approach can dramatically reduce the delay, and consequently, increase the network throughput. The initial researches on network coding were related to the wired networks, but it was gradually generalized to the wireless networks \cite{13,14}. The network codes were first considered in the network layer with the assumption of an ideal and error-free physical layer, but afterwards were extended to the physical layer by considering the effects of fading and noise. The most spectrally efficient Physical Layer Network Coding (PLNC) was proposed in \cite{15,16,17}, in which the nodes simultaneously transmit their own data, and the other nodes receive the linear combinations of the data transmitted plus the noise during just one time slot; nonetheless, a high level of synchronization must be hold in order this scheme to be viable.

Recently, the network coding has been exploited in the cooperative relay networks due to its capability to increase the network throughput. Quite a few of these works have somehow demonstrated that utilizing the linear network coding in the multi-source cooperative networks instead of the traditional schemes leads to the same diversity order, while improves the network throughput. \cite{18} has investigated the network coding in a double-source cooperative network, where each node creates a linear combination of its own symbol and its partner's correctly decoded symbol on $GF(2)$. In \cite{20}, a cooperative transmission scheme based on the network coding along with the multi-user detection has been proposed to improve the users' Bit Error Rate (BER). \cite{21} has introduced a cooperative scheme based on the network coding and the best relay selection technique in a network with $N$ source-destination pairs and $M$ intermediate relays. In \cite{21}, each destination must correctly decode the other sources' symbols in order to recover its own symbol, and without this assumption the diversity order decreases from $M+1$ to $2$. The aforementioned papers have used the binary field $GF(2)$ to build the linear combinations. Nevertheless, some papers considered using the higher fields $GF(q)$. Specifically, \cite{23} showed that in the multi-source networks with the error-prone source-relay channels, the network coding on $GF(2)$ can not lead to the full diversity order. Furthermore, \cite{25} considered a network with $M$ sources where each source acts as a relay for the other sources, and a $(M^2,M)$ maximum distance separable (MDS) linear block code is utilized by the relays. The scheme of \cite{25} leads to the network throughput equal to $1/M$ symbol per channel use (spcu) and diversity order of $2M-1$. Although based on the Singleton bound \cite{34}, the diversity order of $M^2-M+1$ was expected in \cite{25}, but due to the error possibility in the inter-source channels this expectation had not been realized. \cite{28} has generalized the idea of \cite{25}, in which a $(\alpha M^2,\alpha M)$ ($\alpha\geq 2$) MDS linear block code is used as the network coding matrix, which leads to the improved Singleton bound $\alpha(M^2-M)+1$ with the same throughput of $1/M$ spcu. Although these schemes have boosted the diversity order, but they are still suffering from low network throughput. To address this problem, \cite{29} has introduced Complex Field Network Coding (CFNC) scheme in a network including $N$ sources, which reaches to the diversity order of $N$, as well as the network throughput of approximately $1$ spcu. In \cite{30}, the CFNC has been also utilized in a network including $N_S$ sources, $N_R$ relays, and one common destination, which reaches to the full diversity order of $N_R+1$ along with the network throughput of $1/2$ symbol per source per channel use (spspcu) (or equivalently $N_S/2$ spcu), while the network throughput of the traditional (AF and DF) and the finite field linear network coding schemes in such a network are respectively equal to $1/(N_R+1)$ spcu and $N_S/(N_S+N_R)$ spcu. The main reason for the higher throughput of the CFNC schemes in \cite{29} and \cite{30} is the utilization of the complex fields instead of the finite fields, in which the symbols of different nodes, similar to the PLNC, are simultaneously transmitted and their linear combinations in the complex field are received in the other nodes; as a result, this scheme possesses highly complicated synchronization concerns of the PLNC. Furthermore, in \cite{26}, a cooperative transmission scheme based on the linear network codes designed over $GF(q)$ has been introduced for improving the Diversity-Multiplexing Trade-off (DMT) in a network consisting of $N$ source-destination pairs, and $M$ intermediate relays, which reaches the full diversity order of $M+1$ as well as the network throughput of $N/(N+M)$ spcu. In \cite{26}, the parity check matrix of a $(N+M,M,N+1)$ MDS linear block code is used as the network coding matrix. However, the field size required for such MDS codes increases exponentially with the values of $N$ and $M$, which can significantly enhance the system complexity.

In this paper, to increase the diversity order without the degradation of the network throughput, we propose a new cooperative transmission scheme called ``Convolutional Network-Coded Cooperation" (CNCC) in a network including $N$ sources, one relay with $M$ antenna, and one common destination. The CNCC scheme uses the generator matrix of a systematic convolutional code designed on $GF(2)$ throughout the network as the network coding matrix. In this scheme, the systematic packets of the sources are directly transmitted from the sources to the destination, which are simultaneously received and decoded by the relay. If the relay correctly decodes all the packets, it will generate the parity packets by using the underlying convolutional code, and finally will transmit the parity packets by its best antenna (the strongest R-D channel) to the destination. In contrast to the linear network coding scheme, in our proposed CNCC scheme, the diversity order can be enhanced without the reduction of the network throughput, by the increase of the underlying convolutional code's constraint length. It will be shown that the proposed scheme improves the network throughput, and its diversity order at the worst scenarios (weak S-R channels) is equal to $M+1$, and at the best scenarios (strong enough S-R channels) can be much greater than $M+1$. In fact, when the relay is located close to the sources, the diversity order can reach to $\mathcal{D}^{\star}$, where $\mathcal{D}^{\star}\geq d_{free}+M-1$, and $d_{free}$ is the free distance of the exploited convolutional code. It should be noted that the free distance of the convolutional code is increased by the constraint length at the expense of the decoder's complexity. In contrast to the linear network coding scheme, in our proposed scheme, the network throughput does not depend on the number of antennas, but on the number of parity bits of the convolutional codes.

The rest of this paper is organized as follows. In Section II, the system model is described. In Section III, the proposed CNCC scheme is introduced. In Section IV, the performance of the CNCC scheme is analyzed in terms of the sources' BER and the diversity order. Four examples for the CNCC scheme are presented in Section V. The numerical results are provided in Section VI. Ultimately, Section VII concludes the paper, and suggests some future works.

\section{System Model}
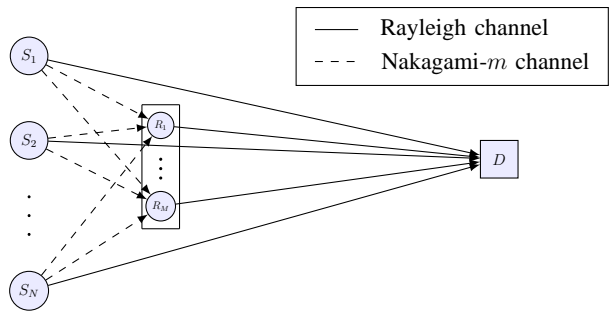
\begin{figure}
\centering
\begin{tikzpicture}[scale=2.5]
\tikzset{>=latex}
\tikzstyle{every node}=[draw,shape=circle,scale=.5,minimum size=1cm,fill=white!92!blue];

\path (0,0) node (v0) {$S_N$};
\path (0,.8) node (v1) {$S_2$};
\path (0,1.25) node (v2) {$S_1$};
\path (.7,.45) node[draw,shape=circle,scale=.7,minimum size=1cm,fill=white!92!blue] (v3) {$R_M$};
\path (.7,.88) node[draw,shape=circle,scale=.7,minimum size=1cm,fill=white!92!blue] (v4) {$R_1$};
\path (2.5,.7) node[rectangle] (v5) {$D$};
\path (0,.3) node[scale=.05,fill=black] (v7){v7};
\path (0,.4) node[scale=.05,fill=black] (v8){v8};
\path (0,.5) node[scale=.05,fill=black] (v9){v9};
\path (.7,.6) node[scale=.05,fill=black] (v10){v10};
\path (.7,.65) node[scale=.05,fill=black] (v11){v11};
\path (.7,.70) node[scale=.05,fill=black] (v12){v12};
\path (2.25,1.3) node[draw,shape=rectangle,scale=1.5,minimum size=1cm,fill=white] (v6) {$\begin{array}{c}\textrm{~~~~~~~~Rayleigh channel~~~~~} \\ \textrm{~~~~~~~~Nakagami-$m$ channel}\end{array}$};
\draw (.6,.33) -- (.8,.33);
\draw (.6,.99) -- (.8,.99);
\draw (.6,.33) -- (.6,.99);
\draw (.8,.33) -- (.8,.99);
\draw (1.52,1.39) -- (1.75,1.39);
\draw [dashed] (1.52,1.22) -- (1.75,1.22);
\draw [->] (v0) -- (v5);
\draw [->] [dashed] (v0) -- (v3);
\draw [->] [dashed] (v0) -- (v4);
\draw [->] (v1) -- (v5);
\draw [->] [dashed] (v1) -- (v3);
\draw [->] [dashed] (v1) -- (v4);
\draw [->] (v2) -- (v5);
\draw [->] [dashed] (v2) -- (v3);
\draw [->] [dashed] (v2) -- (v4);
\draw [->] (v3) -- (v5);
\draw [->] (v4) -- (v5);
\end{tikzpicture}
\captionsetup{font=small,labelfont=footnotesize}
\caption{The cooperative network with $N$ sources $(S_1,...,S_N)$, one relay with $M$ antennas $(R_1,...,R_M)$, and one common destination $(D)$.}

\label{figure-1}
\end{figure}

We consider a cooperative network consisting of $N$ single-antenna sources, one $M$-antenna relay, and one common single-antenna destination, as shown in Fig. \ref{figure-1}. The relay's $M$ antennas are omnidirectional which can be used for both the reception and the transmission. All the S-D, R-D, and S-R channels are assumed to be block fading channels with the depth of $n$ bit transmission intervals, which change independently and identically from one block to another block. Each source has $L$ information bits for transmission which is divided to the $l$ packets of length $n$ bits, where $L=nl$. We assume a perfect interleaving process throughout the network. That is, the packets in the sources and the relay are interleaved before transmissions, and the received packets at the relay and the destination are deinterleaved before decoding. The interleavers must have a sufficient depth (usually greater than $n$) such that the successive bits of a packet experience almost independent fading coefficients. Furthermore, the Binary Phase Shift Keying (BPSK) modulation is used for the transmissions.

The S-D and R-D channels are assumed to be Rayleigh fading channels. Due to the fact that the relay is close to the sources, the S-R channels are assumed to be Nakagami-$m$ fading channels which for $m=1$ are reduced to the Rayleigh, and for $m>1$ act stronger than Rayleigh channels, such that for $m\rightarrow\infty$ Nakagami-$m$ channel converts to the AWGN channel.

All the transmissions are accomplished through orthogonal time slots like the Time Division Multiple Access (TDMA) protocol. The duration of each time slot for transmission of a packet (with the length of $n$ bits) is $T$ second. By using a proper Cyclic Redundancy Check (CRC) code, we assume that the relay is able to detect the erroneous decoded packets. The relay receives the packet of each source, during the source's dedicated time slots, via its $M$ antenna, and then decodes the packet using the Maximum Ratio Combining (MRC) of the $M$ received signals.

\section{The Proposed CNCC Scheme}
The proposed CNCC scheme exploits a good systematic $(N+M^{'}, N, \nu)$ convolutional code designed over $GF(2)$ with the generator matrix of
\begin{equation}
\label{eq1}
G(D)=\left[~I_{N\times N}~ \big\vert~ P_{N\times M^{'}}(D)~\right]
\end{equation}
as the network coding matrix implemented in the network level, where $I_{N\times N}$ is the $N\times N$ identity matrix, and $ P_{N\times M^{'}}(D)$ is an $N\times M^{'}$ matrix whose entries are either a polynomial or a rational function of $D$. Furthermore, $M^{'}$ and $\nu$ are respectively the number of the parity bits and the constraint length of the convolutional code.

The encoder of the convolutional code $G(D)$ in (1) is minimally realized in the relay. That is, the relay contains $\nu$ memories (shift registers). The $N$ systematic packets, related to the first section of the $G(D)$, \textit{i.e.}, $I_{N\times N}$, are directly transmitted from the $N$ sources to the destination within the first $N$ time slots. The $M^{'}$ parity packets are generated from the $N$ correctly decoded sources' packets in the relay, pertaining to the second section of the $G(D)$, \textit{i.e.}, $P_{N\times M^{'}}(D)$, and then are transmitted from the best antennas of the relay to the destination during the consequent $M^{'}$ time slots. The best antenna of the relay at each time slot is defined as the antenna that possesses the strongest R-D channel, which is recognized by the destination.

Specifically, the transmission strategy in the CNCC scheme is as follows. The sources transmit their own interleaved packets of the length $n$ during their dedicated time slots to the destination, where the relay simultaneously receives them through its $M$ antenna, and after deinterleaving the packets, decodes each packet by the MRC method. If the relay correctly decodes all the $i$-th $(i\in\lbrace1,...,l\rbrace)$ $N$ sources' packets (success $(s)$ situation), it will produce the corresponding $M^{'}$ parity packets by using the convolutional code $G(D)$ in (1), and after interleaving, will transmit them from its best antennas to the destination during the dedicated time slots. However, if the relay fails to correctly decode all the $N$ packets (failure $(f)$ situation), it will not generate any parity packets, and will inform the destination. In the failure situations, the sources' packets are merely decoded based on the signals received through the direct S-D paths without the help of the relay. But, in the success situations, the destination uses both the sources' systematic packets and the parity packets received from the relay, and after deinterleaving, runs the Viterbi algorithm to decode all the packets of the sources.

Due to the proximity of the relay to the sources, the number of the failure situations is negligible compared with the number of the success situations. Hence, the network throughput in the proposed CNCC scheme is tightly lower bounded as $\mathcal{R}\geq N/(N+M^{'})$ spcu (symbol per channel use). We are interested in the lower value for $M^{'}$ to increase the network throughput. Accordingly, by selecting the number of parity outputs of convolutional codes less than (or equal to) the number of the relay's antennas $(M^{'}\leq M)$, the network throughput of the CNCC scheme ($N/(N+M^{'})$ spcu) will be greater than (or equal to) that of the LNC scheme with $M$ single-antenna relays ($N/(N+M)$ spcu). That is, the network throughput of the CNCC scheme is not a function of $M$, and consequently, it remains constant and does not decrease with the increase of the number of antennas.

\section{Performance Analysis}

In this section, we first analyze the the BER of the proposed scheme, and then determine the achieved diversity order. As mentioned previously, there is two situations ($s$ and $f$) in the relay that must be considered in the BER analysis. Hence, the end-to-end BER of the network's sources can be written as
\begin{equation}
\label{eq2}
P_b=P_{\left.b\right|s}P_s+P_{\left.b\right|f}P_f=P_{\left.b\right|s}\left(1-P_f\right)+P_{\left.b\right|f}P_f
\end{equation} 
where $P_b$ is the BER of the sources. $P_f$ is the probability of the failure situation in which the relay fails to correctly decode all the $N$ packets of the $N$ sources. $P_s$ is the probability of the success situation in which the relay correctly decodes all the $N$ sources' packets, where $P_s+P_f=1$. $P_{\left.b\right|f}$ and $P_{\left.b\right|s}$ are respectively the BER of the sources in the failure and the success situations.

\subsection{Computation of $P_f$}

The $t$-th received signal from the $i$-th source ($s_i$) at the $j$-th antenna of the relay ($r_j$) is as
\begin{equation}
\label{eq3}
y_{s_i,r_j}\left(t\right)=\sqrt{E_b}\ h_{s_i,r_j}\left(t\right)\ x_{s_i}\left(t\right)+n_{s_i,r_j}\left(t\right)
\end{equation}
where $i=1,...,N$, $j=1,...,M$, and $t=1,...,L$. $h_{s_i,r_j}(t)$s are Nakagami-$m$ fading coefficients from the $s_i$ to the $r_j$. $x_{s_i}\left(t\right)$ is the  BPSK signal transmitted from the $s_i$. Moreover, $n_{s_i,r_j}\left(t\right)$ is the additive white Gaussian noise with the zero mean and the variance of $N_0/2$. $E_b$ is the transmitted energy per bit. We assume that all the S-R channels have the same average energy; that is, $\mathcal{E}\left\{ h^2_{s_i,r_j}\right\}=\overline{h^2_{sr}},~\forall i\in \left\{1,\dots ,N\right\},~\forall j\in \left\{1,\dots ,M\right\}$. Hence, the probability density functions (pdf) of the coefficients are as
\begin{equation}
\label{eq4}
f_{h_{s_i,r_j}}\left(h_{s_i,r_j}\right)=2{\left(\frac{m}{\overline{h^2_{sr}}}\right)}^m\frac{{h_{s_i,r_j}}^{2m-1}}{\Gamma \left(m\right)}\ e^{-m \frac{{h_{s_i,r_j}}^2}{\overline{h^2_{sr}}}},
\end{equation}
where $\Gamma \left(m\right)=\int^{\infty }_0{x^{m-1}e^{-x}dx}$ is the Gamma function, and for integer values of $m$ is equal to $\Gamma \left(m\right)=(m-1)!$.

First, we calculate the bit error probability $P_e$ at the relay. By using the MRC, the conditional bit error probability of a BPSK signal is as

\begin{equation}
\label{eq5}
P_{e|{\gamma }_{s_i,r}}=Q\left(\sqrt{2{\gamma }_{s_i,r}}\right),
\end{equation}

\begin{equation}
\label{eq6}
{\gamma }_{s_i,r}=\frac{E_b}{N_0}\sum^M_{j=1}{h^2_{s_i,r_j}},
\end{equation}
where ${\gamma }_{s_i,r}$ is the received SNR of the transmitted signal from the source $s_i$ at the relay. By defining $h^{'}_{s_i,r_j}=h^2_{s_i,r_j}$, and due to the fact that $h_{s_i,r_j}$ is the Nakagami-$m$ random variable, given in \eqref{eq4}, $h^{'}_{s_i,r_j}$ will be a Chi Square random variable with $2m$ degrees of freedom as
\begin{equation}
\label{eq7}
f_{h^{'}_{s_i,r_j}}\left(h^{'}_{s_i,r_j}\right)={\left(\frac{m}{\overline{h^2_{sr}}}\right)}^m\frac{{h^{'}_{s_i,r_j}}^{m-1}}{\Gamma \left(m\right)}\ e^{-m\ \frac{h^{'}_{s_i,r_j}}{\overline{h^2_{sr}}}}
\end{equation}
$ \forall i\in \left\{1,\dots,N\right\},~\forall j\in \left\{1,\dots ,M\right\} $. Hence, $\gamma_{s_i,r}$ in \eqref{eq6}, which is the sum of $M$ independent Chi Square
random variables each of which with $2m$ degrees of freedom, has the Chi Square pdf with $2Mm$ degrees of freedom. By defining $\overline{\gamma}_{sr}=\frac{E_b}{N_0}\overline{h^2_{sr}}$ as the average received SNR of the S-R channels, we have
\begin{equation}
\label{eq11}
f_{\gamma_{s_i,r}} \left(\gamma_{s_i,r}\right)={\left(\frac{m}{\overline{\gamma}_{sr}}\right)}^{Mm}\frac{{\gamma_{s_i,r}}^{Mm-1}}{\Gamma \left(Mm\right)}\ e^{-m\frac{\gamma_{s_i,r}}{\overline{\gamma}_{sr}}}.
\end{equation}
As a result, the unconditional bit error probability, $P_e$, can be easily obtained from \eqref{eq5} and \eqref{eq11} as
\begin{equation}
\label{eq12}
P_e=\int^{\infty}_{\gamma_{s_i,r}=0} P_{e|{\gamma }_{s_i,r}}f_{\gamma_{s_i,r}} \left(\gamma_{s_i,r}\right)d\gamma_{s_i,r},
\end{equation}
\begin{equation}
\label{eq13}
P_e=\left[\frac{1}{2}(1-\mu_{sr})\right]^{Mm}\sum^{Mm-1}_{w=0}{Mm-1+w \choose w }\left[\frac{1}{2}(1+\mu_{sr})\right]^w,
\end{equation}
where $\mu_{sr}=\sqrt{\frac{\overline{\gamma}_{sr}}{m+\overline{\gamma}_{sr}}}$.

Now, we compute the failure probability, $P_f$. Because of assuming a perfect interleaving, the successive bits within each sources' packets experience independent fadings. As a result, the probability that one packet of a specific source can be correctly decoded in the relay is equal to $(1-P_e)^n$. In \eqref{eq2}, $P_s$ is the probability that all the $N$ packets of the $N$ sources corresponding to the $N$ successive slots can be correctly decoded in the relay. Therefore, the success probability at the relay is as follows
\begin{equation}
\label{eq14}
P_s=(1-P_e)^{Nn}.
\end{equation}
As a result, $P_f$ in \eqref{eq2} is computed as
\begin{equation}
\label{eq15}
P_f=1-P_s=1-(1-P_e)^{Nn},
\end{equation}
where $P_e$ is given by \eqref{eq13}.

At the high SNRs ($\overline{\gamma}_{sr}\rightarrow\infty$), $P_e$ and $P_f$ can be respectively approximated as
\begin{equation}
\label{eq16}
P_e\approx {2Mm-1 \choose Mm}\left(\frac{m}{4\overline{\gamma}_{sr}}\right)^{Mm},
\end{equation}

\begin{equation}
\label{eq17}
P_f\approx nNP_e\approx K\left(\frac{1}{\overline{\gamma}_{sr}}\right)^{Mm},
\end{equation}
where $\footnotesize K=nN{2Mm-1 \choose Mm}\left(\frac{m}{4}\right)^{Mm}$ is a constant coefficient.

\subsection{Computation of $P_{\left.b\right|f}$}
When the relay fails to correctly decode the $N$ packets of the sources, it does not participate in the cooperation phase, and consequently, these packets are decoded only based on the received signals through the direct S-D channels. Hence, the bit error probability of the sources in the failure situation is simply obtained similar to \eqref{eq13} by setting $m=1$ and $M=1$, and substituting $\overline{\gamma}_{sd}$ instead of  $\overline{\gamma}_{sr}$. As a result, we have
\begin{equation}
\label{eq18}
P_{\left.b\right|f}=\frac{1}{2}\left(1-\sqrt{\frac{\overline{\gamma}_{sd}}{1+\overline{\gamma}_{sd}}}\right),
\end{equation}
where $\overline{\gamma}_{sd}$ is defined as the average received SNR from the S-D channels at the destination. Similar to \eqref{eq16}, at the high SNRs ($\overline{\gamma}_{sd}\rightarrow\infty$), $P_{\left.b\right|f}$ can be approximated as
\begin{equation}
\label{eq19}
P_{\left.b\right|f}\approx\frac{1}{4\overline{\gamma}_{sd}}.
\end{equation}

\subsection{Computation of $P_{\left.b\right|s}$}
In the $s$ situation, based on the $N$ packets of the sources, the relay produces the corresponding $M^{'}$ parity packets using the systematic $(N+M^{'}, N, \nu)$ convolutional code given in \eqref{eq1}. Finally, the parity packets are transmitted through the best antenna of the relay during their dedicated $M^{'}$ time slots. The destination runs the Viterbi algorithm to decode the sources' packets. Hence, the BER of the sources in this situation is equal to the BER of the exploited convolutional code described by $G(D)$ in \eqref{eq1} whose systematic and parity packets are respectively transmitted through the Rayleigh fading S-D channels, and the best of $M$ available Rayleigh fading R-D channels.

Due to interleaving with sufficient depth, the successive bits of each packet sent by the sources and the relay are well assumed to experience independent fadings. The received signals from the $N$ sources and the best selected antennas of the relay in the destination are respectively  as follows
\begin{equation}
\label{eq20}
y_{s_i,d}(t)=\sqrt{E_b}h_{s_i,d}(t)x_{s_i,d}(t)+z_{s_i,d}(t),
\vspace{-.7cm}
\end{equation}
\begin{equation}
\label{eq21}
y_{r_{sel},d}(t^{'})=\sqrt{E_b}h_{r_{sel},d}(t^{'})x_{r_{sel},d}(t^{'})+z_{r_{sel},d}(t^{'}),
\end{equation}
where $i \in \lbrace 1,...,N\rbrace$, $t \in \lbrace 1,...,L\rbrace$, and $t^{'} \in \lbrace 1,...,M^{'}L\rbrace$. The parameters in \eqref{eq20} and \eqref{eq21} are as follows. $E_b$: the transmitted energy per bit. $x_{s_i,d}(t)$: the $t$-th transmitted bit from the $i$-th source. $h_{s_i,d}(t)$ and $z_{s_i,d}(t)$: respectively, the Rayleigh fading coefficient, and the Gaussian noise with zero mean and the variance of $N_0/2$. $x_{r_{sel},d}(t^{'})$: the $t^{'}$-th parity bit transmitted from the best selected antenna of the relay. $h_{r_{sel},d}(t^{'})$ and $z_{r_{sel},d}(t^{'})$: respectively, the Rayleigh fading coefficient, and the Gaussian noise with zero mean and the variance of $N_0/2$.

The BER of a $(N+M^{'},N,\nu)$ convolutional code with the free distance of $d_{free}$ is upper bounded as
\begin{equation}
\label{eq22}
P_{b_{conv.}}<\frac{1}{N}\sum^\infty_{d=d_{free}}B_dP_d,
\end{equation}
where $B_d$s are the coefficients of the Bit Weight Enumerating Function (BWEF) of the convolutional code as
\begin{equation}
\label{eq23}
B(X)=\sum^\infty_{d=d_{free}}B_dX^d=\frac{\partial A(W,X)}{\partial W}\big|_{W=1}.
\end{equation}
$A(W,X)$ is the Input-Output Weight Enumeration Function (IOWEF) of the code, which can be easily computed from the state diagram of the convolutional code. Furthermore, $P_d$ in \eqref{eq22} is the Pairwise Error Probability (PEP) with the Hamming weight of $d$. It must be noticed that $P_d$ only depends on the S-D and R-D channels. 

\subsubsection{Computation of PEP}
The destination uses the Maximum Likelihood criterion to decode the sequence transmitted by the sources as well as the best antennas of the relay. Hence, according to \eqref{eq20} and \eqref{eq21}, the conditional PEP can be easily obtained as \cite{37}
\begin{equation}
\label{eq24}
P_d|_{\lbrace \gamma_{sd},\gamma_{r_{sel},d}\rbrace}=Q\left(\sqrt{2\sum^{d_1}_{k=1} \gamma_{sd}(t_k)+2\sum^{d_2}_{k=1}\gamma_{r_{sel},d}(t^{'}_k)}\right),
\end{equation}
where $d_1+d_2=d$. $t_k$ and $t^{'}_k$ are the error positions related to the transmitted bits respectively from the sources and the relay. Moreover, $\gamma_{sd}(t_k)=\frac{E_b}{N_0}h^2_{sd}(t_k)$ and $\gamma_{r_{sel},d}(t^{'}_k)=\frac{E_b}{N_0}h^2_{r_{sel},d}(t^{'}_k)$ are, respectively, the instantaneous received SNRs from the sources and the best antennas of the relay. The $\gamma_{sd}(t_k)$s are independent for different $k$, and have the exponential pdf as
\begin{equation}
\label{eq25}
f_{\gamma_{sd}(t_k)}\left(\gamma_{sd}(t_k)\right)=\frac{1}{\overline{\gamma}_{sd}}\exp \frac{-\gamma_{sd}(t_k)}{\overline{\gamma}_{sd}},
\end{equation}
where $\overline{\gamma}_{sd}=\frac{E_b}{N_0}\overline{h^2_{sd}}$ is the average received SNR from the S-D channels. Furthermore, $\gamma_{r_{sel},d}(t^{'}_k)$ is as
\begin{equation}
\label{eq26}
\gamma_{r_{sel},d}(t^{'}_k)=\max_{j=1,...,M}\gamma_{r_j,d}(t^{'}_k),
\end{equation}
where $\gamma_{r_j,d}(t^{'}_k)=\frac{E_b}{N_0}h^2_{r_j,d}(t^{'}_k)$ is the received SNR from the $j$-th antenna of the relay at the destination, and has the exponential pdf and cdf, respectively, as  
\begin{equation}
\label{eq27}
f_{\gamma_{r_j,d}(t^{'}_k)}\left(\gamma_{r_j,d}(t^{'}_k)\right)=\frac{1}{\overline{\gamma}_{rd}}\exp \frac{-\gamma_{r_j,d}(t^{'}_k)}{\overline{\gamma}_{rd}}
\vspace{-.5cm}
\end{equation}
\begin{equation}
\label{eq28}
F_{\gamma_{r_j,d}(t^{'}_k)}\left(\gamma_{r_j,d}(t^{'}_k)\right)=1-\exp \frac{-\gamma_{r_j,d}(t^{'}_k)}{\overline{\gamma}_{rd}}
\end{equation}
$\forall j \in \lbrace 1,...,M \rbrace$, where $\overline{\gamma}_{rd}=\frac{E_b}{N_0}\overline{h^2_{rd}}$ is the average received SNR from the R-D channels. Then, from \eqref{eq26}-\eqref{eq28}, the pdf of $\gamma_{r_{sel},d}(t^{'}_k)$ is easily obtained as

\begin{equation} 
\label{eq29}
f_{\gamma_{r_{sel},d}(t^{'}_k)}\left(\gamma_{r_{sel},d}(t^{'}_k)\right)=M\left[1-\exp \frac{-\gamma_{r_{sel},d}(t^{'}_k)}{\overline{\gamma}_{rd}}\right]^{M-1}\frac{1}{\overline{\gamma}_{rd}}\exp \frac{-\gamma_{r_{sel},d}(t^{'}_k)}{\overline{\gamma}_{rd}}
\end{equation}

Now, we can compute $P_d$ by averaging $ P_d|_{\lbrace \gamma_{sd},\gamma_{r_{sel},d}\rbrace} $ in \eqref{eq24} over the distributions of $  \gamma_{sd}(t_k) $ and $ \gamma_{r_{sel},d} (t^{'}_k) $ given respectively in \eqref{eq25} and \eqref{eq29} as

\begin{eqnarray}
\label{eq30}
P_d=\int^\infty_0 ... \int^\infty_0 Q\left(\sqrt{2\sum^{d_1}_{k=1} \gamma_{sd}(t_k)+2\sum^{d_2}_{k=1}\gamma_{r_{sel},d}(t^{'}_k)}\right) \left(\prod^{d_1}_{k=1}\frac{1}{\overline{\gamma}_{sd}}\exp \frac{-\gamma_{sd}(t_k)}{\overline{\gamma}_{sd}}\right)\times~~~~~~~~~~~      \nonumber\\
\left(\prod^{d_2}_{k=1}M\left[1-\exp \frac{-\gamma_{r_{sel},d}(t^{'}_k)}{\overline{\gamma}_{rd}}\right]^{M-1}\frac{1}{\overline{\gamma}_{rd}}\exp \frac{-\gamma_{r_{sel},d}(t^{'}_k)}{\overline{\gamma}_{rd}}\right)\prod^{d_1}_{k=1}d\gamma_{sd}(t_k)\prod^{d_2}_{k=1}d\gamma_{r_{sel},d}(t^{'}_k).
\end{eqnarray} 
By using the upper bound  $ Q(x)\leq \frac{1}{2} e^{\frac{-x^2}{2}}, ~x\geq0 $  and the Binomial expansion, $(1+x)^n=\sum^n_{w=0}{n\choose w}x^w$, and after some straightforward simplifications, the upper bound for $P_d$ is obtained as

\begin{equation}
\label{eq33}
P_d\leq\frac{1}{2}\left(\frac{1}{1+\overline{\gamma}_{sd}}\right)^{d_1}\left(M\sum^{M-1}_{w=0}{M-1 \choose w}(-1)^w\frac{1}{1+w+\overline{\gamma}_{rd}}\right)^{d_2},
\end{equation}
where $d_1+d_2=d$. It can be easily demonstrated that the following inequality holds

\begin{equation}
\label{eq34}
M\sum^{M-1}_{w=0}{M-1 \choose w}(-1)^w\frac{1}{1+w+\overline{\gamma}_{rd}}<\frac{M(M-1)!}{\overline{\gamma}_{rd}^{~~~M}}.
\end{equation}
Hence, from \eqref{eq33} and \eqref{eq34}, we also have

\begin{equation}
\label{eq35}
P_d\leq\frac{1}{2}\left(\frac{1}{1+\overline{\gamma}_{sd}}\right)^{d_1}\left(\frac{M(M-1)!}{\overline{\gamma}_{rd}^{~~~M}}\right)^{d_2},
\end{equation}
where $d_1+d_2=d$.

From \eqref{eq33} and \eqref{eq35}, $P_d$ is not an explicit function of $d$, but a function of $d_1$ and $d_2$ such that $d_1+d_2=d$. Hence, we can not directly use the equations \eqref{eq22} and \eqref{eq23} to compute the BER of the sources in the success situations, because the given BWEF is an explicit function of $d$. For the problem on hand, we define a modified IOWEF, $A^{Mod}(W,Y,Z)$, which is the function of $d_1$ and $d_2$ as follows
\begin{equation}
\label{eq36}
A^{Mod}(W,Y,Z)=\sum_{w,d_1,d_2}A_{w,d_1,d_2}W^wY^{d_1}Z^{d_2},
\end{equation}
where $d_1+d_2=d$ and $d\geq d_{free}$. In this equation, the exponents of $W$, $Y$, and $Z$ denote respectively the Hamming weights of the input bits, the systematic output bits, and the parity output bits of the convolutional code. Furthermore, $ A_{w,d_1,d_2} $ is the number of paths in the state diagram of the code, which originate from the zero state, and finally return to the zero state, such that their numbers of nonzero input bits, nonzero systematic output bits, and nonzero parity output bits are respectively equal to $w$, $d_1$, and $d_2$. Since the exploited convolutional code in the CNCC scheme is systematic, we always have $w=d_1$. To derive $A^{Mod}(W,Y,Z)$, in the state diagram of the exploited code $G(D)$ in \eqref{eq1}, we assign a gain $W^{\alpha_b}Y^{\beta_b}Z^{\zeta_b}$ to each branch $b$, such that $\alpha_b$, $\beta_b$, and $\zeta_b$ are respectively the Hamming weights of the input block (among $N$ input bits), the systematic output block (among the first $N$ output bits), and the parity output block (among the last $M^{'}$ output bits) of the branch $b$. Again, we have $\alpha_b=\beta_b$. With this dedicated gain to each branch, the transfer function from the initial zero state to the final zero state in the modified state diagram yields the $A^{Mod}(W,Y,Z)$. Similar to \eqref{eq23}, the modified BWEF is obtained from the modified IOWEF as
\begin{equation}
\label{eq37}
B^{Mod}(Y,Z)=\sum_{d_1,d_2}B_{d_1,d_2}Y^{d_1}Z^{d_2}=\frac{\partial A^{Mod}(W,Y,Z)}{\partial W}\big|_{W=1},
\end{equation}
where $d_1+d_2=d$ and $d\geq d_{free}$. $B_{d_1,d_2}$ is the total number of the nonzero input bits in the paths of the modified state diagram with the number of the nonzero systematic output bits equal to $d_1$, and the number of the nonzero parity output bits equal to $d_2$.

Now, similar to \eqref{eq22}, the BER of the sources in the success situations can be expressed as
\begin{equation}
\label{eq38}
P_{\left.b\right|s}<\frac{1}{N}\sum_{d_1,d_2}B_{d_1,d_2}P_d.
\end{equation}
Ultimately, from \eqref{eq33} and \eqref{eq35}-\eqref{eq38}, the closed form upper bounds for the $P_{\left.b\right|s}$ are obtained as
\begin{equation}
\label{eq41}
P_{\left.b\right|s}<\frac{1}{2N}B^{Mod}\left(Y=\frac{1}{1+\overline{\gamma}_{sd}},~Z=M\sum^{M-1}_{w=0}{M-1 \choose w}(-1)^w\frac{1}{1+w+\overline{\gamma}_{rd}}\right),
\end{equation}
and
\begin{equation}
\label{eq42}
P_{\left.b\right|s}<\frac{1}{2N}B^{Mod}\left(Y=\frac{1}{1+\overline{\gamma}_{sd}},~Z=\frac{M(M-1)!}{\overline{\gamma}_{rd}^{~~~M}}\right).
\end{equation}

\subsection{End-to-end BER and the achieved diversity order}
Taking the path loss effect into account leads to the following relationships: $\overline{\gamma}_{sr}=\left(\frac{d_{sd}}{d_{sr}}\right)^\eta \overline{\gamma}_{sd}$ and $\overline{\gamma}_{rd}=\left(\frac{d_{sd}}{d_{rd}}\right)^\eta \overline{\gamma}_{sd}$, where $d_{sr}$, $d_{sd}$, and $d_{rd}$ denote respectively the S-R, S-D, and R-D distances, and $\eta$ is the path loss exponent. Without loss of generality, we assume that the relation $d_{sd}=d_{sr}+d_{rd}$ approximately holds. Therefore, we have  $\overline{\gamma}_{sr}=\beta^\eta \overline{\gamma}_{sd}$ and $\overline{\gamma}_{rd}=\left(\frac{\beta}{\beta-1}\right)^\eta \overline{\gamma}_{sd}$, where $\beta=\frac{d_{sd}}{d_{sr}}$. We define $\overline{\gamma}\triangleq \overline{\gamma}_{sd}$ as the average received SNR in the destination. Hence, from \eqref{eq2}, \eqref{eq13}, \eqref{eq15}, \eqref{eq18}, and \eqref{eq41}, the upper bound of the end-to-end BER of the network's sources is obtained as

\begin{eqnarray}
\label{eq43}
P_b<\frac{1}{2N}B^{Mod}\left(Y=\frac{1}{1+\overline{\gamma}},~Z=M\sum^{M-1}_{w=0}{M-1 \choose w}(-1)^w\frac{1}{1+w+\overline{\gamma}_{rd}}\right)\times    \nonumber\\
\left(1-\left[\frac{1}{2}(1-\mu_{sr})\right]^{Mm}\sum^{Mm-1}_{w=0}{Mm-1+w \choose w }\left[\frac{1}{2}(1+\mu_{sr})\right]^w\right)^{nN} \nonumber\\
+\frac{1}{2}\left(1-\sqrt{\frac{\overline{\gamma}}{1+\overline{\gamma}}}\right)\times \left(1-\left(1-\left[\frac{1}{2}(1-\mu_{sr})\right]^{Mm}\sum^{Mm-1}_{w=0}{Mm-1+w \choose w }\left[\frac{1}{2}(1+\mu_{sr})\right]^w\right)^{nN}\right)
\end{eqnarray}
where $\mu_{sr}=\sqrt{\frac{\overline{\gamma}_{sr}}{m+\overline{\gamma}_{sr}}}$, $\overline{\gamma}_{sr}=\beta^\eta \overline{\gamma}$, $\overline{\gamma}_{rd}=\left(\frac{\beta}{\beta-1}\right)^\eta \overline{\gamma}$, and $\beta=\frac{d_{sd}}{d_{sr}}$.

Now, we aim to compute the diversity order of the CNCC scheme as
\begin{equation}
\label{eq44}
\mathcal{D}\triangleq -\lim_{\overline{\gamma}\rightarrow \infty}\frac{\log P_b(\overline{\gamma})}{\log \overline{\gamma}}.
\end{equation}
To this end, we first analyze the achieved diversity order in the success situation from \eqref{eq42} and \eqref{eq37}. Let's introduce the set $\mathcal{F}$ as
\begin{equation}
\label{eq45}
\mathcal{F}=\left\lbrace~(d_1,d_2)~\big|~Y^{d_1}Z^{d_2}~\mathrm{exists~in~the~series~expression~of}~B^{Mod}(Y,Z) ~\right\rbrace.
\end{equation}
We then define
\begin{equation}
\label{eq46}
(d_1^{\star},d_2^{\star})=\arg \min_{(d_1,d_2)} d_1+Md_2  ~~~\mathrm{s.t.}~~~ (d_1,d_2)\in \mathcal{F}.
\end{equation}
It would be worthy to mention that the pair $(d_1^{\star},d_2^{\star})$ is not necessarily unique. Then, from \eqref{eq37}, \eqref{eq42}, and \eqref{eq44}-\eqref{eq46}, it can be easily observed that the diversity order in the success situation is equal to
\begin{equation}
\label{eq47}
\mathcal{D}^{\star}=d_1^{\star}+Md_2^{\star},
\end{equation}
which can be rewritten as
\begin{equation}
\label{eq50}
\mathcal{D}^{\star}=d_1^{\star}+d_2^{\star}+(M-1)d_2^{\star}.
\end{equation}
Due to the fact that $d_1^{\star}$ and $d_2^{\star}$ are respectively related to the error patterns of the sources and the relay, $d_1^{\star}\geq1$ and $d_2^{\star}\geq1$. Moreover, we have $d_1^{\star}+d_2^{\star}\geq d_{free}$. Hence, from \eqref{eq50}, $ \mathcal{D}^{\star} $ can be lower bounded as
\begin{equation}
\label{eq51}
\mathcal{D}^{\star}\geq d_{free}+M-1.
\end{equation}

Finally, from \eqref{eq1}, \eqref{eq17}, \eqref{eq19}, \eqref{eq42}, \eqref{eq45}, and \eqref{eq46}, at high SNRs ($\overline{\gamma}\rightarrow\infty$), $P_b$ is approximated as
\begin{equation}
\label{eq48}
P_b\approx K^{'}\left(\frac{1}{\overline{\gamma}}\right)^{d_1^{\star}}\left(\frac{M(M-1)!}{\overline{\gamma}^M}\right)^{d_2^{\star}}+K^{''}\left(\frac{1}{\overline{\gamma}}\right)^{Mm+1}
\end{equation}
where $K^{'}$ and $K^{''}$ are the constant terms. According to \eqref{eq44}, \eqref{eq47}, and \eqref{eq48}, the achieved diversity order of the sources in the CNCC scheme is equal to 
\begin{equation}
\label{eq49}
\mathcal{D}_{\mathrm{CNCC}}=\min ~(\mathcal{D}^{\star}, Mm+1).
\end{equation}
By a proper value of $\nu$ in the exploited $(N+M^{'},N,\nu)$ convolutional code, we can always have $d_{free}\geq M+1$, and consequently, $\mathcal{D}^{\star} \geq M+1$. We consider two special cases.
\subsubsection{Weak S-R channels}
In this case, we consider the Rayleigh fading channels (Nakagami-$m$ with $m=1$), the same as the S-D and the R-D channels. Hence, the diversity order will be $\mathcal{D}_{\mathrm{CNCC}}=M+1$.
\subsubsection{Strong S-R channels}
For this case which is more likely to happen, according to the assumption that the relay is close to the sources, the Nakagami-$m$ S-R channels with $m>1$ are well assumed. As a result, the diversity order up to $\mathcal{D}_{\mathrm{CNCC}}=\mathcal{D}^{\star}$ can be achieved, which is much more than $M+1$, as will be shown in the examples of the next section.


\section{Illustrative Examples For The CNCC Scheme}
In this section, we consider four networks with two sources, and different number of the relay's antennas and $M^{'}$. For all the examples considered, the constraint lengths of the exploited convolutional codes have been set to $\nu=3$.

\subsection{First ($N=2, M=1, M^{'}=1$) and second ($N=2, M=2, M^{'}=1$) networks}
\begin{figure}
\centering
\tikzset{XOR/.style={draw,circle,append after command={
        [shorten >=\pgflinewidth, shorten <=\pgflinewidth,]
        (\tikzlastnode.north) edge (\tikzlastnode.south)
        (\tikzlastnode.east) edge (\tikzlastnode.west)
        }
    }
}
\tikzset{line/.style={draw, -latex',shorten <=1bp,shorten >=1bp}}

\begin{tikzpicture}[scale=1.1]
\tikzset{>=latex}
\node (A) at (6.5,1.05) [rectangle, draw=white,scale=.8] {$V^{(1)}$};
\node (A) at (6.5,.6) [rectangle, draw=white,scale=.8] {$V^{(2)}$};
\node (A) at (6.5,0) [rectangle, draw=white,scale=.8] {$V^{(3)}$};
\node (A) at (6.5,1.05) [rectangle, draw=white,scale=.8] {$V^{(1)}$};
\node (A) at (-1,1.05) [rectangle, draw=white,scale=.8] {$U^{(1)}$};
\node (A) at (-1,.6) [rectangle, draw=white,scale=.8] {$U^{(2)}$};

\tikzstyle{every node}=[draw,shape=circle,scale=.5,minimum size=1cm,fill=white!92!blue];

\path (0,0) node (XOR-aa)[XOR,scale=.4,fill=white] (1) {};

\path (1.8,0) node (XOR-aa)[XOR,scale=.4,fill=white] (2) {};

\path (3.6,0) node (XOR-aa)[XOR,scale=.4,fill=white] (3) {};

\path (5.4,0) node (XOR-aa)[XOR,scale=.4,fill=white] (4) {};

\path (.9,0) node[rectangle,minimum width=1.8cm,minimum height=.8cm] (5) {};

\path (2.7,0) node[rectangle,minimum width=1.8cm,minimum height=.8cm] (6) {};
\path (4.5,0) node[rectangle,minimum width=1.8cm,minimum height=.8cm] (7) {};

\path (0,1.05) node[circle, scale=.15, fill=black] {};
\path (-.4,.6) node[circle, scale=.15, fill=black] {};
\path (1.8,1.05) node[circle, scale=.15, fill=black] {};
\path (1.6,.6) node[circle, scale=.15, fill=black] {};
\path (3.6,1.05) node[circle, scale=.15, fill=black] {};
\path (3.6,-.6) node[circle, scale=.15, fill=black] {};
\path (5.4,1.05) node[circle, scale=.15, fill=black] {};
\path (5.2,.6) node[circle, scale=.15, fill=black] {};
\path (5.7,0) node[circle, scale=.15, fill=black] {};

\draw [->] (-.4,0) -- (1);
\draw [->] (1) -- (5);
\draw [->] (5) -- (2);
\draw [->] (2) -- (6);
\draw [->] (6) -- (3);
\draw [->] (3) -- (7);
\draw [->] (7) -- (4);
\draw [->] (4) -- (6.1,0);
\draw [->] (-.6,.6) -- (6.1,.6);
\draw [->] (-.6,1.05) -- (6.1,1.05);
\draw (-.4,.6)--(-.4,0);
\draw (5.7,0)--(5.7,-.6);
\draw (5.7,-.6)--(0,-.6);
\draw [->] (0,-.6) -- (0,-.1);
\draw [->] (0,1.05) -- (0,.1);
\draw [->] (1.8,1.05) -- (1.8,.1);
\draw [->] (3.6,1.05) -- (3.6,.1);
\draw [->] (5.4,1.05) -- (5.4,.1);
\draw [->] (3.6,-.6) -- (3.6,-.1);
\draw (1.6,.6)--(1.6,.2);
\draw [->] (1.6,.2)--(1.73,.07);
\draw (5.2,.6)--(5.2,.2);
\draw [->] (5.2,.2)--(5.33,.07);
\end{tikzpicture}
\captionsetup{font=small,labelfont=footnotesize}
\caption{The encoder of the convolutional code $G_1(D)$ realized in the observer canonical form.}
\vspace{+.2cm}
\label{figure-2}
\end{figure}
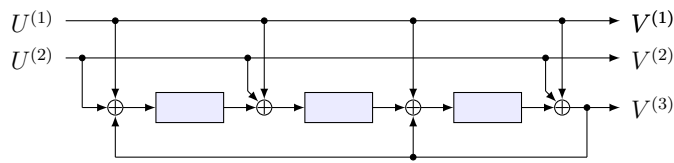

We select $M^{'}=1$ in these networks. Hence, the CNCC scheme exploits the good systematic $(3,2,3)$ convolutional code as
\begin{equation}
\label{eq52}
G_1(D)=
\begin{bmatrix}
1&0&\frac{1+D+D^2+D^3}{1+D+D^3}\\
0&1&\frac{1+D^2+D^3}{1+D+D^3}
\end{bmatrix}
\end{equation}
with $d_{free}=4$. The encoder of this code can be minimally realized in the relay by three memories, as shown in Fig. \ref{figure-2}, where $U^{(1)}$ and $U^{(2)}$ are the input bits, $V^{(1)}$ and $V^{(2)}$ are the systematic output bits, and $V^{(3)}$ is the parity output bit generated and transmitted by the relay. The modified BWEF of $G_1(D)$ has been computed by Matlab. Its first several terms which play the significant roles in the performance of the CNCC scheme is as follows
\begin{eqnarray}
\label{eq53}
B_1^{Mod}(Y,Z)=(3Y^3Z)+(4Y^4Z+9Y^3Z^2+2Y^2Z^3) \nonumber\\
+(44Y^4Z^2+18Y^3Z^3)+\dots
\end{eqnarray}
The sum of the exponents of $Y$ and $Z$ in \eqref{eq53} represents the related Hamming weights of the code sequences, which the least value is in fact the free distance of the code. According to \eqref{eq45}-\eqref{eq47} and \eqref{eq53}, for both the first and the second networks we have $(d_1^{\star},d_2^{\star})=(3,1)$ (related to the first term of \eqref{eq53}) which leads to $\mathcal{D}^{\star}=d_1^{\star}+d_2^{\star}=4$ and $\mathcal{D}^{\star}=d_1^{\star}+2d_2^{\star}=5$ for the first and the second networks, respectively.

\subsection{Third ($N=2, M=2, M^{'}=2$) and fourth ($N=2, M=3, M^{'}=2$) networks}
We select $M^{'}=2$ in these networks. Hence, the CNCC scheme exploits the good systematic $(4,2,3)$ convolutional code as
\begin{equation}
\label{eq54}
G_2(D)=
\begin{bmatrix}
1&0&\frac{1+D^2+D^3}{1+D+D^3}&\frac{1+D^2}{1+D+D^3}\\
0&1&\frac{D^2}{1+D+D^3}&\frac{1+D^2+D^3}{1+D+D^3}
\end{bmatrix}
\end{equation}
with $d_{free}=6$. The modified BWEF of $G_2(D)$ has been obtained as
\begin{eqnarray}
\label{eq57}
B_2^{Mod}(Y,Z)=(3Y^3Z^3)+(5Y^5Z^2+12Y^4Z^3+9Y^3Z^4) \nonumber\\
+(10Y^5Z^3+20Y^4Z^4)+\dots
\end{eqnarray}
From \eqref{eq57}, we have $(d_1^{\star},d_2^{\star})=(3,3)$ and $(5,2)$ (related to the first and the second terms of \eqref{eq57}), and as a result, $\mathcal{D}^{\star}=d_1^{\star}+2d_2^{\star}=9$ for the third network. For the fourth network, we have $(d_1^{\star},d_2^{\star})=(5,2)$, and consequently, $\mathcal{D}^{\star}=d_1^{\star}+3d_2^{\star}=11$.

As can be realized from these examples, the diversity order is much higher than $M+1$. Furthermore, by the increase of the constraint length, the diversity order can be further enhanced. The achieved diversity orders and the network throughputs of these four networks have been given in Table I for different values of $m$. For the comparisons, the diversity order and the network throughput of the linear network coding scheme with $N$ users and $M$ relays (each with one antenna) have been included as well. As expected, the proposed CNCC scheme outperforms the LNC (as a result, the conventional cooperative schemes as well) in terms of the diversity order as well as the network throughput for the examples considered at the expense of the receiver complexity at the destination.

\begin{table}[!t]
\captionsetup{font=small,labelfont=footnotesize}
\caption{The achieved diversity orders and the network throughputs in the four example networks}
\vspace{-.2cm}
\label{table-1}
\centering
\begin{tabular}{|c|c|c|c|c|c|}
\hline
Network & $M^{'}$ & \hspace{-.2cm} $\begin{array}{c}\textrm{CNCC} \hspace{-.1cm} \\ \textrm{with}\end{array}$ & $\mathcal{D^{\star}}$ & $\begin{array}{c}\textrm{Diversity order} \\ \textrm{(CNCC, LNC)}  \\ \begin{tabular}{cccc} \hline \hspace{-.3cm}  $m=1$&$m=2$&$m=3$&$m=4$ \hspace{-.3cm} \end{tabular}  \end{array}$ & \hspace{-.3cm} $\begin{array}{c}\textrm{Network }  \hspace{-.2cm}  \\   \textrm{throughput~(spcu)} \hspace{-.2cm}  \\ \begin{tabular}{cc} \hline \hspace{-.3cm}   CNCC~&~~LNC  \hspace{-.3cm} \end{tabular}     \end{array}$ \\
\hline
$1\textrm{st:} ~N=2, M=1$ & $1$ & $ G_1(D) $ & $ 4 $&$(2,2)~~~~(3,2)~~~~(4,2)~~~~(4,2)$ & $2/3~~~~~~~2/3$\\
\hline
$2\textrm{nd:} ~N=2, M=2$ & $1$ & $ G_1(D) $ & $ 5 $&$(3,3)~~~~(5,3)~~~~(5,3)~~~~(5,3)$ & $2/3~~~~~~~1/2$\\
\hline
$3\textrm{rd:} ~N=2, M=2$ & $2$ & $ G_2(D) $ & $ 9 $&$(3,3)~~~~(5,3)~~~~(7,3)~~~~(9,3)$ & $1/2~~~~~~~1/2$\\
\hline
$4\textrm{th:} ~N=2, M=3$ & $2$ & $ G_2(D) $ & $ 11 $&$~(4,4)~~~~(7,4)~~~(10,4)~~~(11,4)$ & $1/2~~~~~~~2/5$\\
\hline
\end{tabular}

\end{table}

\section{Numerical Results}
\begin{figure}
\centering
\includegraphics[width =3.5in, keepaspectratio]{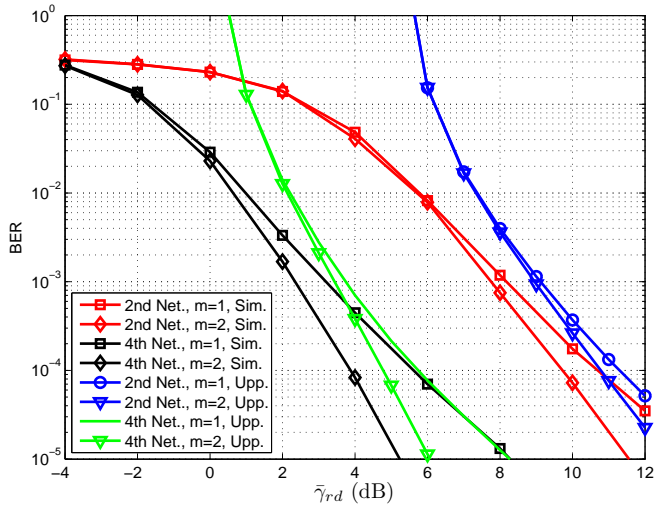}
\captionsetup{font=small,labelfont=footnotesize}
\caption{BER vs. $\overline{\gamma}_{rd}$ in the CNCC scheme with $n=10$ and $\frac{d_{sd}}{d_{sr}}=3$. Simulation results and upper bound \eqref{eq43}.}

\label{figure-3}
\end{figure}

\begin{figure}
\centering
\includegraphics[width =3.5in, keepaspectratio]{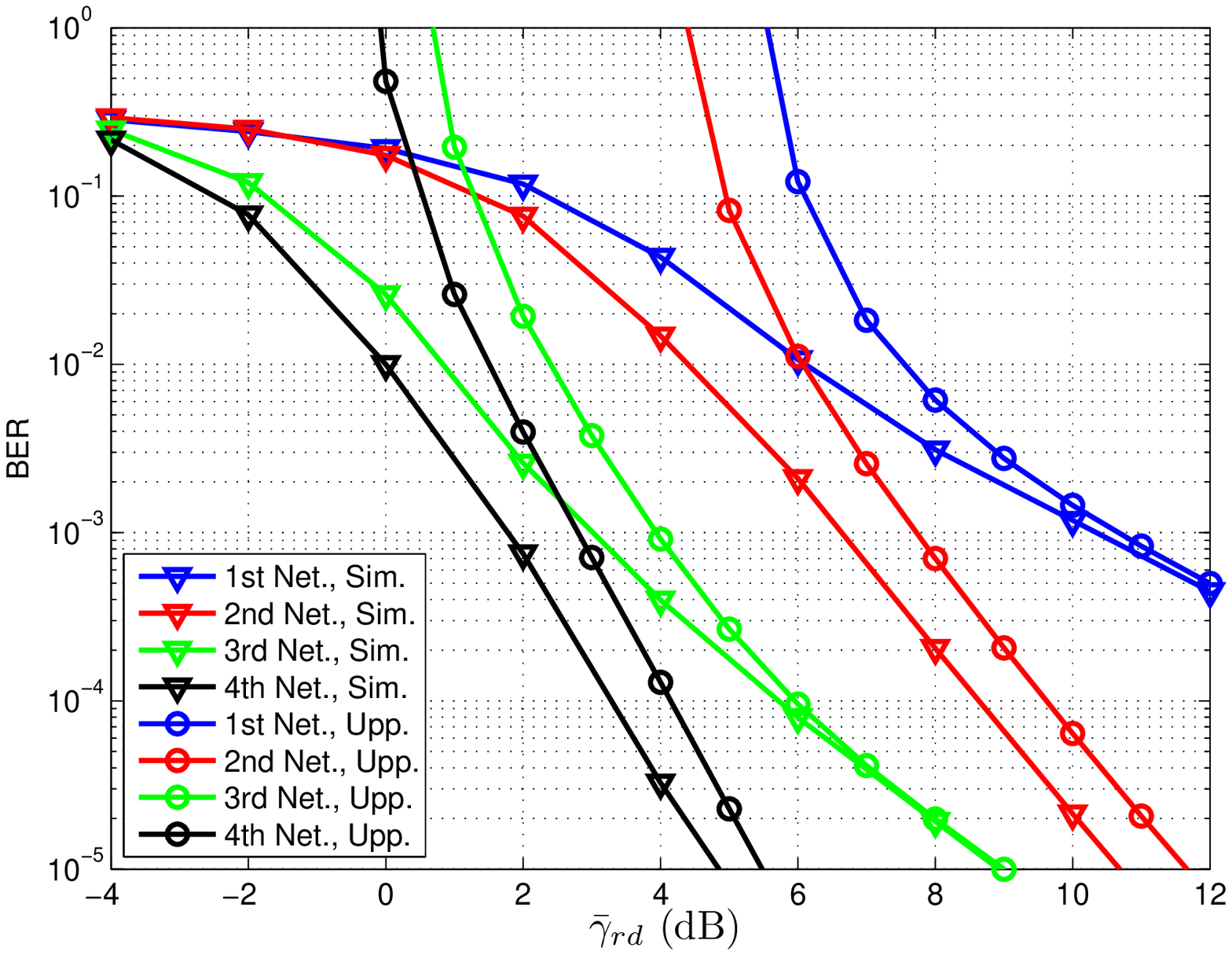}
\captionsetup{font=small,labelfont=footnotesize}
\caption{BER vs. $\overline{\gamma}_{rd}$ in the CNCC scheme with $m=1$, $n=10$ and $\frac{d_{sd}}{d_{sr}}=5$. Simulation results and upper bound \eqref{eq43}.}

\label{figure-4}
\end{figure}

\begin{figure}
\centering
\includegraphics[width =3.5in, keepaspectratio]{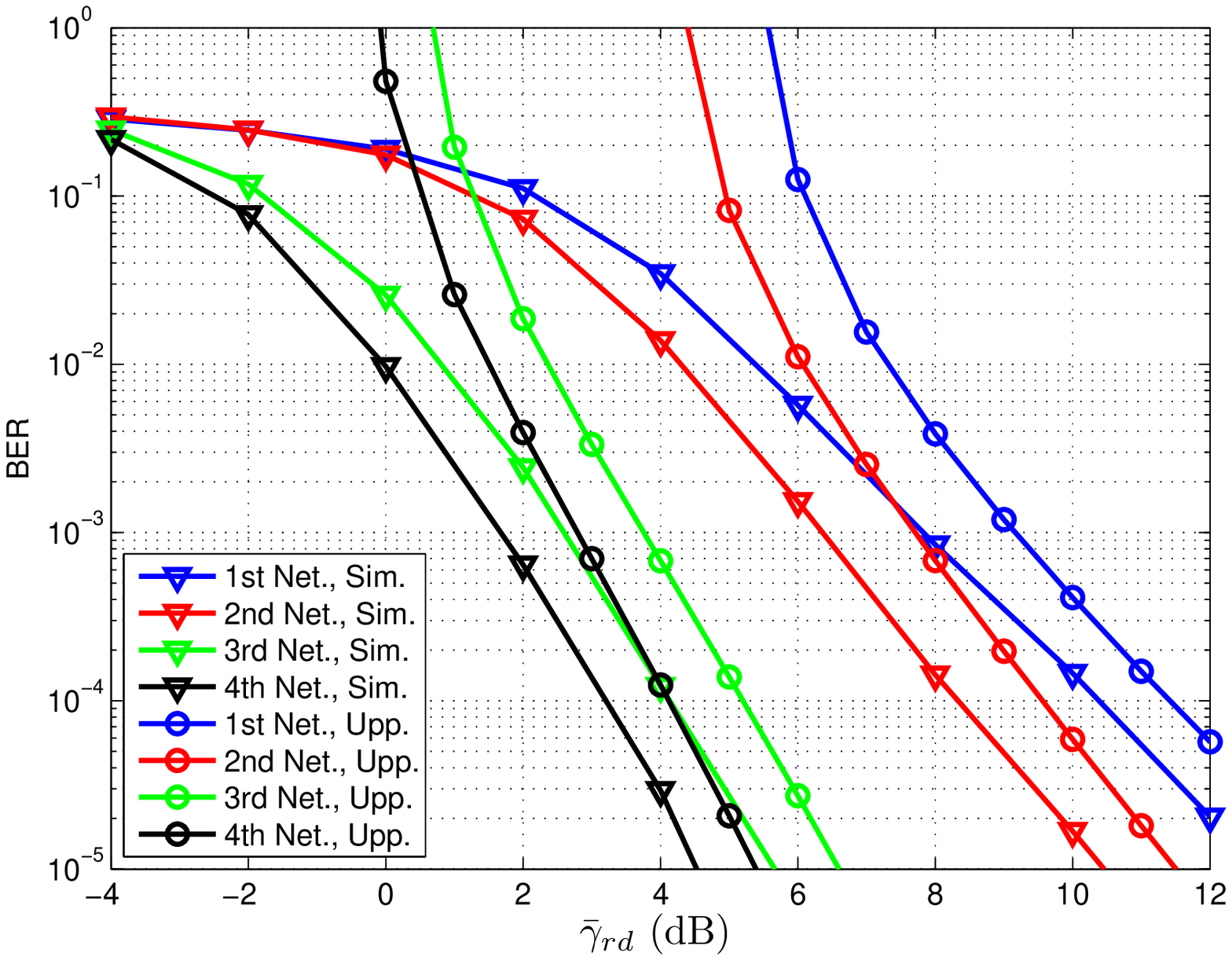}
\captionsetup{font=small,labelfont=footnotesize}
\caption{BER vs. $\overline{\gamma}_{rd}$ in the CNCC scheme with $m=2$, $n=10$ and $\frac{d_{sd}}{d_{sr}}=5$. Simulation results and upper bound \eqref{eq43}.}

\label{figure-5}
\end{figure}

\begin{figure}
\centering
\includegraphics[width =3.5in, keepaspectratio]{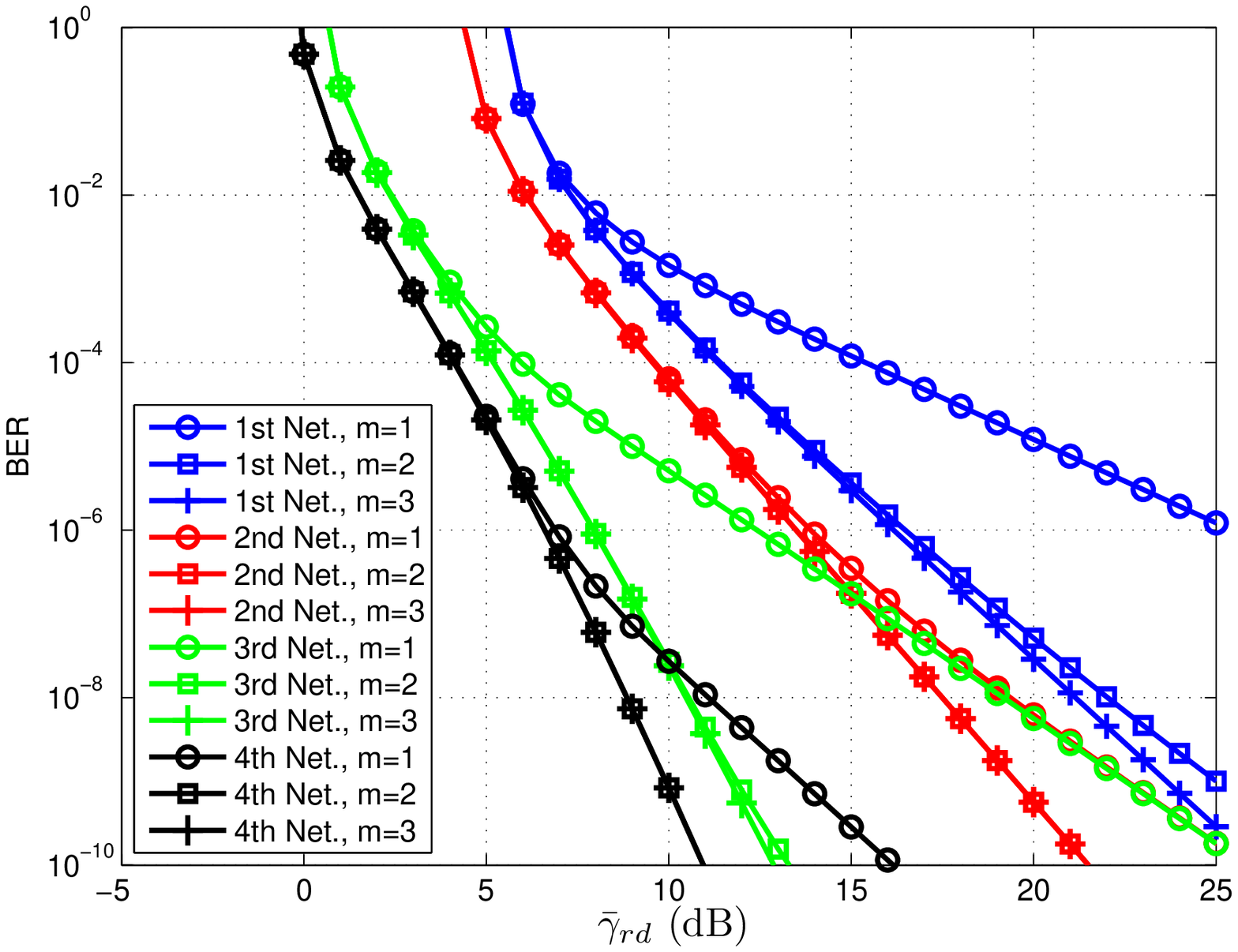}
\captionsetup{font=small,labelfont=footnotesize}
\caption{BER vs. $\overline{\gamma}_{rd}$ in the CNCC scheme with $n=10$ and $\frac{d_{sd}}{d_{sr}}=5$. Upper bound \eqref{eq43}.}

\label{figure-6}
\end{figure}

\begin{figure}
\centering
\includegraphics[width =3.5in, keepaspectratio]{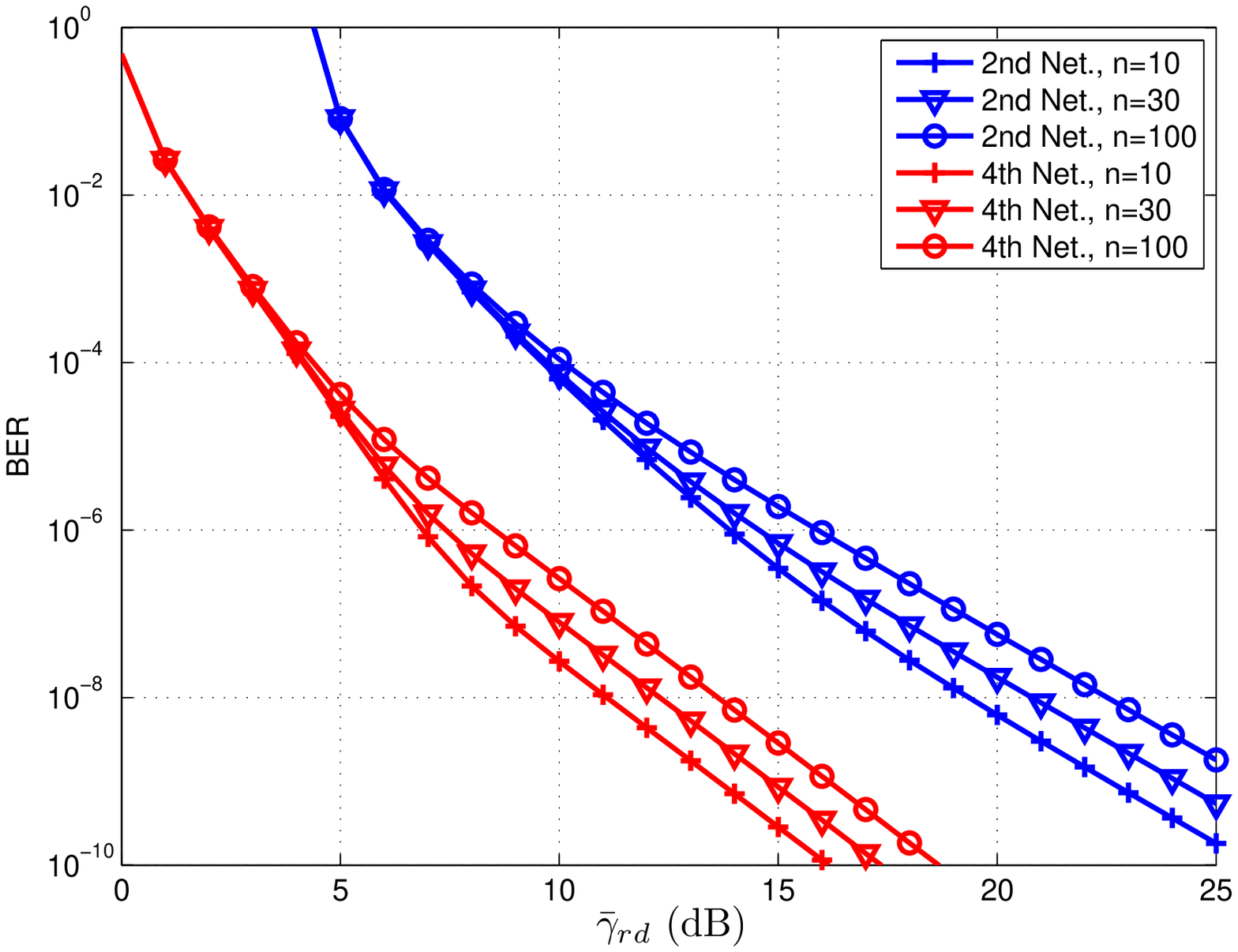}

\captionsetup{font=small,labelfont=footnotesize}
\caption{BER vs. $\overline{\gamma}_{rd}$ in the CNCC scheme with $m=1$ and $\frac{d_{sd}}{d_{sr}}=5$. Upper bound \eqref{eq43}.}
\label{figure-9}
\end{figure}

We have provided some numerical results to evaluate the performance of our proposed scheme considering the networks of the previous section. In all the simulations, except in Fig. \ref{figure-9}, we set $n=10$, $\eta=2$ (related to the free space), and interleaving depth of $100$. Figs. \ref{figure-3}, \ref{figure-4}, and \ref{figure-5} represent the plots of the sources' BERs versus $\overline{\gamma}_{rd}$ (average received SNR of the R-D channel) along with the upper bound \eqref{eq43} for the different scenarios considering $\frac{d_{sd}}{d_{sr}}=3$ and $\frac{d_{sd}}{d_{sr}}=5$. Fig. \ref{figure-3} indicates the results of the second and the fourth networks for $m=1$ (Rayleigh fading) and $m=2$ considering $\frac{d_{sd}}{d_{sr}}=3$. Furthermore, the results of the four networks, considering $\frac{d_{sd}}{d_{sr}}=5$, have been given in Figs. \ref{figure-4} and \ref{figure-5} respectively for $m=1$ and $m=2$. As can be seen, the upper bound \eqref{eq43} is consistent with the numerical results in all figures. It can be concluded that the increase of $m$ leads to the BER improvement especially at high SNRs, but at low SNRs, the BERs are approximately the same. According to Figs. \ref{figure-3} and \ref{figure-4}, as the relay's position becomes closer to the sources in the cases that $M>M^{'}$, the CNCC scheme performs more stable and robust against the weak S-R channel conditions; that is, when $M>M^{'}$ (second and fourth networks) and the relay is close enough to the sources, the BER results are identical for $m=1$ and $m=2$ at the practical ranges of SNR. As expected, from Fig. \ref{figure-4}, which is related to the Rayleigh S-R channels, the network does not perform quite well for $M=1$, compared with the other examples with $M>1$. In all figures, the slopes of the BER plots are exactly the same as those of the upper bounds, which indicate the achieved diversity orders. For further investigating the achievable diversity orders, in Fig. \ref{figure-6} the upper bound \eqref{eq43} has been depicted for $m=1$, $m=2$, and $m=3$ at the high SNRs. In this figure, the diversity orders match with the theoretical results given in \eqref{eq49}. Fig. \ref{figure-6} demonstrates that the most considerable degradation of the BERs occurs when the S-R channels are Rayleigh fading ($m=1$), but the BER results are approximately identical for $m\geq 2$. Moreover, according to Fig. \ref{figure-6}, when $M^{'}<M$ (second and fourth networks), the BERs are less vulnerable to the weak S-R channel conditions ($m=1$) than when $M^{'}=M$ (first and third networks). Both the second and the fourth networks present the same BERs for $m=1$, $m=2$, and $m=3$ at the BER range of up to $10^{-6}$, while for the first and the third networks, the performance in the Rayleigh S-R channels ($m=1$) are much worse than those of $m=2$ and $m=3$.

\begin{figure}
\centering
\includegraphics[width =3.5in, keepaspectratio]{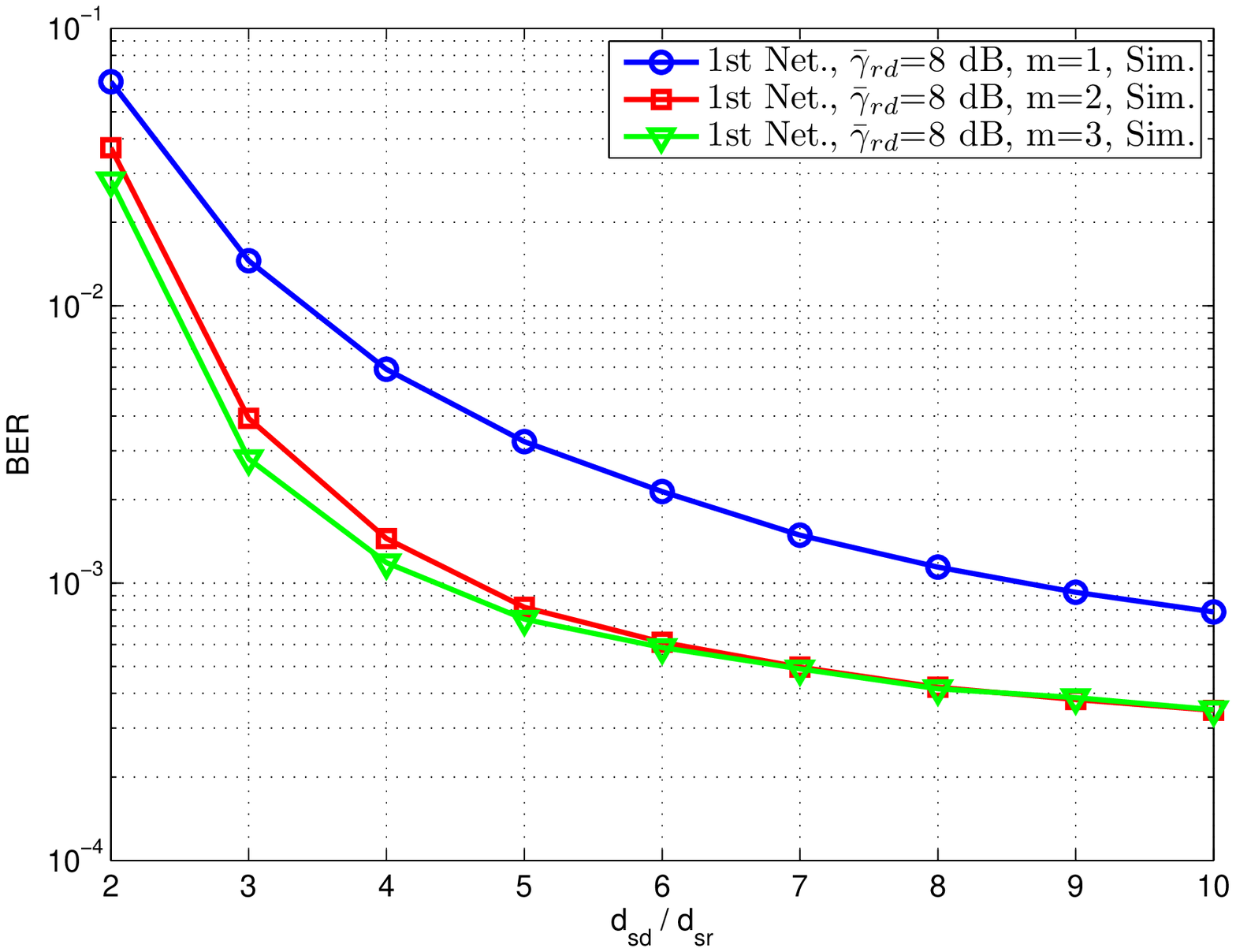}
\captionsetup{font=small,labelfont=footnotesize}
\caption{BER vs. $\frac{d_{sd}}{d_{sr}}$ in the CNCC scheme with $n=10$ at $\overline{\gamma}_{rd}=8$ dB. Simulation results.}

\label{figure-7}
\end{figure}

\begin{figure}
\centering
\includegraphics[width =3.5in, keepaspectratio]{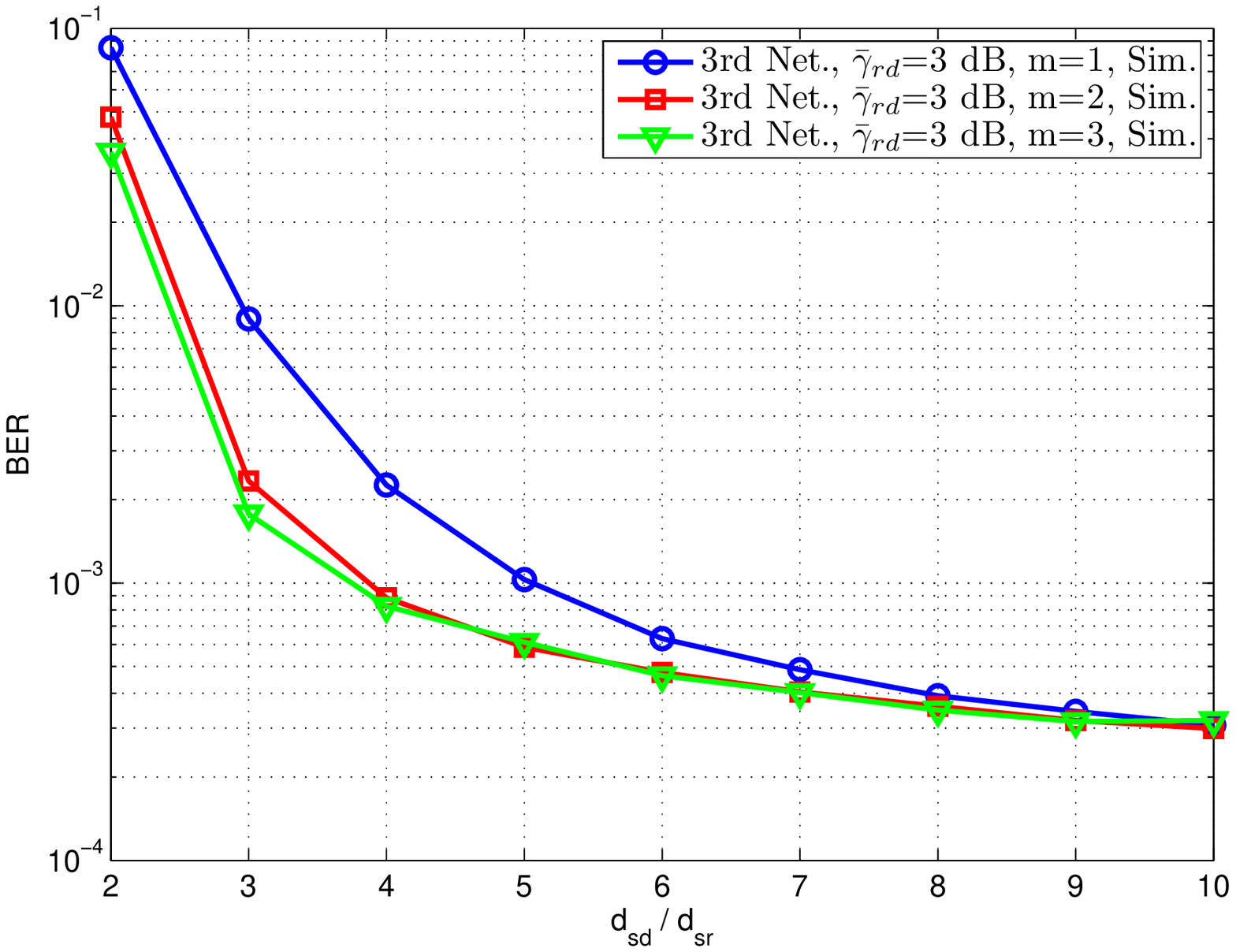}
\captionsetup{font=small,labelfont=footnotesize}
\caption{BER vs. $\frac{d_{sd}}{d_{sr}}$ in the CNCC scheme with $n=10$ at $\overline{\gamma}_{rd}=3$ dB. Simulation results.}

\label{figure-8}
\end{figure}

The effect of the relay's position, between the sources and the destination, on the proposed scheme's performance has been scrutinized in Figs. \ref{figure-7} and \ref{figure-8}, respectively for the first and the third networks. More specifically, the plots of the BERs have been depicted versus $\frac{d_{sd}}{d_{sr}}$ for $m=1$, $m=2$, and $m=3$ at the given SNRs of $\overline{\gamma}_{rd}=8$ dB for the first network, and $\overline{\gamma}_{rd}=3$ dB for the third one. It can be deduced from Figs. \ref{figure-7} and \ref{figure-8} that as the relay locates closer to the sources (that is, $\frac{d_{sd}}{d_{sr}}$ increases), the BERs of the sources decrease. Again, we see from these figures that the BER results for $m=2$ and $m=3$ are nearly the same, and the BER for $m=1$ is worse than them, especially for the first network with one antenna. Based on Figs. \ref{figure-7} and \ref{figure-8}, by the increase of $\frac{d_{sd}}{d_{sr}}$, the sources' BERs for different $m$ converge to the same values which are related to the ideal S-R channels. The convergence rate depends on the number of the relay's antennas ($M$) and the S-R channel conditions ($m$), where the least is related to the first network ($M=1$) with Rayleigh S-R channels ($m=1$) as shown in Fig. \ref{figure-7}.

Finally, Fig. \ref{figure-9} investigates the impact of the transmitted packet lengths ($n$) on the sources' upper bound \eqref{eq43}. In this figure, the upper bounds have been depicted for $n=10,30,100$ in both the second and the fourth networks considering $m=1$ and $\frac{d_{sd}}{d_{sr}}=5$. According to Fig. \ref{figure-9}, the increase of the packet length can slightly deteriorate the BER without reducing the achieved diversity order. However, at the low SNRs (BERs up to $10^{-4}$), the BER degradation is negligible.

\section{Conclusions and Future Works}
In this paper, we proposed a new cooperative transmission scheme called ``CNCC" in the multi-source networks including one multi-antenna relay. The upper bound of the BER of the CNCC scheme was evaluated and then accordingly the achieved diversity order was computed. In a network with $N$ sources and one $M$-antenna relay, the CNCC scheme exploits a good systematic $(N+M^{'}, N, \nu)$ convolutional code over $GF(2)$ as the network coding matrix which is run at the network level. It was realized that the proposed CNCC scheme can simultaneously enhance the network throughput as well as the diversity order compared to the traditional AF and DF, and the LNC-based cooperative schemes. This is because the CNCC's network throughput is approximately equal to $N/(N+M^{'})$ spcu, which for $M^{'}\leq M$ is greater than (or equal to) the network throughputs of LNC ($N/(N+M)$ spcu) and traditional ($1/(1+M)$ spcu) schemes with $M$ single-antenna relays. Furthermore, it was demonstrated that although the diversity order of the CNCC at the worst scenario (Rayleigh S-R channels) reduces to $M+1$ which is equal to the diversity order of the LNC and the traditional schemes, but the CNCC can have better performance due to using the convolutional codes. However, at the most practical scenarios (strong S-R channels), the diversity order of the CNCC scheme reaches to $\mathcal{D}^{\star}$ which can be much more than $M+1$ ($\mathcal{D}^{\star}\geq d_{free}+M-1$ where $d_{free}$ is the free distance of the used convolutional code) by the increase of the underlying convolutional code's constraint length. In addition, the provided simulation results for the four considered examples verified the accuracy of the theoretical analysis. 

It has been realized that the failure situation in which the relay fails to correctly decode all the sources' packets restricts the diversity order. As a suggestion to overcome this problem, in the failure situation, instead of not cooperating, the relay can simply amplify the maximum ratio combined signal corresponding to each packet, and transmit it to the destination, which can lead to the improvement of the diversity order in the weak S-R channel conditions.

\ifCLASSOPTIONcaptionsoff
  \newpage
\fi

\end{document}